\documentclass{article}
\usepackage{graphicx}
\usepackage{epsfig}
\usepackage{amsfonts}
\usepackage{amsmath, amssymb, array}
\usepackage{graphics,graphicx,float,mfpic}
\usepackage{epsfig}
\usepackage{color}
\newcommand{\be}{\begin{equation}}
\newcommand{\ee}{\end{equation}}
\newcommand{\bea}{\begin{eqnarray}}
\newcommand{\eea}{\end{eqnarray}}
\newcommand{\bmat}{\begin{pmatrix}}
\newcommand{\emat}{\end{pmatrix}}

\title{Perturbative non uniform black strings in $\mbox{AdS}_6$}
\author{{\large T\'erence Delsate }\\ \\{\small Physique-Math\'ematique, Universite de Mons-Hainaut, Mons, Belgium}}
\date{\today}

\begin{document}
\maketitle
\begin{abstract}
We construct the non uniform AdS black string solution with a perturbation theory in six dimensions, focusing on the backreacting second order correction. The backreactions at second order give the first relevant corrections to the thermodynamical quantities. 
Our results show that for configurations with horizon radius and length in the extradimension small compared to the AdS radius,
the properties of the non uniform black string are similar to the locally asymptotically flat case. For black strings with small horizon radial coordinate but large length in the extradimension,
the thermodynamical properties of the solutions are affected by the AdS curvature.
\end{abstract}

\section{Introduction}

The last decade has witnessed a growing interest for solutions in AdS space, largely motivated by the AdS/CFT correspondence \cite{adscft2,adscft}. This conjecture states that a solution of Einstein field equations which approaches asymptotically the AdS space is dual to a conformal field theory living on the conformal boundary of the AdS. In this context, black objects are of great interest; for example, the Hawking-Page phase transition \cite{hptr} between the five dimensional spherically symmetric black holes and the thermal AdS background was interpreted by Witten, through AdS/CFT, as a thermal phase transition from a confining to a deconfining phase in the dual four dimensional $\mathcal N = 4$ super Yang-Mills theory \cite{whp}. 

In four dimensional asymptotically flat space, there is a uniqueness theorem on black holes, implying that the horizon topology of black objects is $S_2$. In higher dimensions, this theorem does not hold and many efforts have been done in order to construct different kinds of black objects, such as black strings \cite{solubs} of horizon topology $S_{d-3}\times S_1$, black rings \cite{solubr} (toroidal horizon topology), black holes ($S_{d-3}$) \cite{solubh} and their generalisation with charge, rotation, cosmological constant, etc.

Gregory and Laflamme \cite{GL} showed in '93 that black strings are unstable towards long wavelength perturbations in asymptotically flat space  (see \cite{glrev} for a review). After that discovery, many efforts have been done to understand the end point of the black string instability in asymptotically locally flat spacetimes. It was believed that the endpoint of the instability should be a caged black hole, where the asymptotical space in $d+1$ dimensions is $\mathcal M_d\times S_1$ (with $\mathcal M_d$ the $d$-dimensional Minkowski space), but it was argued \cite{Horowitz:2001cz} that a transition between a black string and a caged black hole would take an infinite proper time at the horizon. This lead to the study of non uniform black string, a black object which is non translational invariant along the $S_1$ coordinate, first in a perturbative approach \cite{gubser} (see also \cite{kudoh}) and after in the full non linear regime \cite{Wiseman:2002zc}. The resulting thermodynamical phase diagram \cite{flatphase} is now well known although dynamical phase transitions are still to be constructed. In this diagram, the non uniform and the caged black hole branch meet at the merger point, where a topological phase transition is expected to occur.

 More recently, it has been shown that this instability persists in asymptotically locally AdS space for small AdS black strings \cite{rbd} where the ratio $r_h/\ell <<1$, with $r_h$ the horizon radius and $\ell$ the AdS radius, fixed by the cosmological constant $\Lambda$. There exists another phase of AdS black strings, namely large AdS black strings, where $r_h/\ell \geq 1$ which is thermodynamically and dynamically stable \cite{rbd}. The picture for these two phases is essentially similar to $AdS$ black hole: small AdS black strings thermodynamically unstable while large black strings are stable. As stated in \cite{hptr}, AdS space acts like a confining box: when the ratio $r_h/\ell$ becomes larger than some critical values, the wavelength of the instable mode cannot fit the 'AdS box' and thus all acceptable wavelength do not lead to instabilities.

However, in asymptotically locally AdS spaces, much less is known about the counterparts of the solutions of $\Lambda = 0$ Kaluza-Klein black objects The evolution of AdS black holes has been studied \cite{gmbh} but still has to be done for AdS black strings. In order to construct a phase diagram in AdS, one has to consider various stationary solutions and compare their thermodynamical properties since these stationary solutions should be equilibrium configuration. This could give some clues on the endpoint of dynamical evolution and phase transitions between different black objects in AdS.

In this paper, we consider the case of perturbative non uniform black strings in $\mbox{AdS}_6$ (sometimes referred as \emph{pNUBS}). We choosed to work in six dimensions because it turns out to be numerically easier than in odd dimension ($log$ terms arise in odd dimensions and are more difficult to deal with) and in more than six dimensions (where the field decay is such that it is much difficult to have an accurate extraction of numerical quantities).  The phase of non uniform black strings is connected to the uniform black string phase, since one can smoothly deform a uniform black string into a non uniform one. The perturbative approach will give the direction of the evolution of the thermodynamical quantities.

This paper is organised as follow: In section \ref{model}, we present and explain the strategy for the perturbative approach. Section \ref{asymptotics} is devoted to the asymptotic solutions for the background, the first order perturbation and the back reaction. We explain in details the numerical technique and the appropriate boundary conditions in section \ref{num} and give the properties of the solutions in section \ref{properties}. Finally, we expose and summarise our results in section \ref{results} and \ref{end}.

\section{The model and the equations}
\label{model}
We consider the Einstein-Hilbert action with a negative cosmological constant in six dimensions, supplemented with Hawking-Gibbons boundary term:
\be
S = \frac{1}{16\pi G}\int_{\mathcal M}d^6x \sqrt{-g}\left( R - \frac{20}{\ell^2}\right) + \frac{1}{8\pi G}\int_{\partial\mathcal M}d^5x \sqrt{-h} K,
\ee
where $G$ is the six dimensional Newton constant, $\ell$ the AdS radius, $R$ the scalar curvature on the manifold $\mathcal M$, $g$ the determinant of the metric on $\mathcal M$, $h$ the determinant of the induced metric on the boundary $\partial \mathcal M$ and $K$ the extrinsic curvature of the boundary.

In order to construct black string solutions, we consider the following ansatz for the metric:
\be
\label{anubs}
ds^2 = -b(r)e^{2A(r,z)}dt^2 + e^{2B(r,z)}\left( \frac{dr^2}{f(r)} + a(r)dz^2 \right) + r^2 e^{2C(r,z)}d\Omega_3^2.
\ee
We furthermore assume the extradimension $z$ to be of finite size, $z\in [0,L]$ and the solution to be periodic in $z$. The ansatz of \cite{rms} corresponds to $A=B=C=0$ and leads to the uniform AdS black strings parameterized by the functions $a,\ b,\ f$ which can be obtained numerically.

With the ansatz \eqref{anubs}, the Einstein equations reduce to a coupled system of partial differential equations. Since it might turn out difficult to solve the equations, we approach the full nonlinear problem with perturbation theory. To do so, we consider $A\ ,B,\ C$ to be $z$-dependent perturbations of the uniform AdS black string. Accordingly, we parameterize the functions according to:
\bea
A(r,z) &=& \epsilon A_1(r)\cos(kz) + \epsilon^2\left[ A_0(r) + A_2(r)\cos(2kz)  \right] + \mathcal O(\epsilon^3),\nonumber\\
B(r,z) &=& \epsilon B_1(r)\cos(kz) + \epsilon^2\left[ B_0(r) + B_2(r)\cos(2kz)  \right] + \mathcal O(\epsilon^3),\\
C(r,z) &=& \epsilon C_1(r)\cos(kz) + \epsilon^2\left[ C_0(r) + C_2(r)\cos(2kz)  \right] + \mathcal O(\epsilon^3),\nonumber
\eea
with $\epsilon<<1$ and $k = 2\pi/L \in\mathbb R$. This small parameter $\epsilon$ is related to the deformation parameter of the string, as we will see later.

The fields $a,b,f$ are the background fields and a solution for the background has been constructed in \cite{rms}. The fields $A_1,B_1,C_1$ are the linear perturbations and give access to the stability problem \cite{rbd}. The fields $A_0,B_0,C_0$ are back reactions, while $A_2,B_2,C_2$ are massive modes (see \cite{gubser} for a construction of these fields in asymptotically flat space). The equations at second order in $\epsilon$ decouple for the massive mode and the back reaction, the massive mode equations come with a $cos$ factor while the back reaction equations do not. 

\par The equations for the background and first order can be found respectively in \cite{rms} and \cite{rbd}. Let us remind the first order equations ($B_1$ can be expressed in terms of $A_1,\ C_1$):
\be
\label{eqF1}
A_1''= \alpha_1 A_1 + \alpha_2 A_1' + \alpha_3 C_1 + \alpha_4 C_1',~~
C_1'' = \varphi_1 A_1 + \varphi_2 A_1' + \varphi_3 C_1 + \varphi_4 C_1',
\ee
where 
\bea
\alpha_1 &=& \frac{2b\left( 3 k^2l^2 -5ra' \right)  + 
    r\left( k^2 \ell^2 + 10 a \right) b'}{\ell^2af
    \left( 6 b + rb' \right) },~~
\nonumber
\label{eqs-p}
\alpha_2 =-\frac{1}{r} - \frac{b'}{2b} - 
  \frac{1}{\ell^2
     rf}\left[5 r^2 + 2 \ell^2  - 
     \frac{20 r^2b}{6b + rb'}\right],
 \\    
     \alpha_3 &=&\frac{30 b\left( 2a - ra' \right) }
  {\ell^2af\left( 6 b + rb' \right) },~~
\alpha_4 =-\frac{\  3 b'   }{2b} + 
  \frac{60 rb}
   {\ell^2 f\left( 6 b + rb' \right) },~~
\nonumber
\varphi_1 =\frac{\left(10 r^2 + 2 \ell^2  \right) 
    \left( -  b a'  + a b' \right) }{\ell^2 r a f
    \left( 6b + r b' \right) },\\
\varphi_2 &=&\frac{1}{r}\left[-1 + \frac{4 b\left( 5 r^2 + 2 \ell^2  \right)
         }{\ell^2 f\left( 6b + rb' \right) }\right],~~
         \nonumber
\varphi_3 =\frac{6b \left( k^2 \ell^2 r + 10 ra - 
       \left( 5r^2 + 2 \ell^2  \right) a'
       \right)  + \ell^2\left( k^2r^2 - 4 a \right) b'}
    {\ell^2 raf\left( 6 b + rb' \right) },
\nonumber
\\
\varphi_4 &=&  \frac{6b \left( k^2 \ell^2r + 10 ra - 
       \left( 5r^2 + 2 \ell^2  \right) a'
       \right)  + \ell^2\left( k^2r^2 - 4 a \right) b'}
    {\ell^2raf\left(6 b + rb' \right) }.
\nonumber
\end{eqnarray}
This is an eigenvalue problem, where $k$ plays the role of the eigenvalue. In fact, the value obtained for $k$ by solving the above equations is the critical Gregory-Laflamme wavenumber for wich the perturbation is static and fixes static perturbative non uniform black string length $L$ since $L = 2\pi/k$.

The equations for the higher order correction follows the pattern of those in asymptotically flat spacetime \cite{gubser}. It is possible to solve $B_0$ in terms of $A_0, C_0$ and of the lower order fields; the equations for the back reacting fields read: 
\bea
\label{eqbr}
A_0'' &+& \frac{1}{2}\left(\frac{4+10\frac{r^2}{\ell^2}+8f}{rf}+ \frac{b'}{b}-\frac{4b(3 + 10\frac{r^2}{\ell^2}+6f)}{rf(6b+rb')}\right) A_0' + \left(\frac{9}{r}+\frac{3b'}{2b}-\frac{6b(3 + 10\frac{r^2}{\ell^2}+6f)}{rf(6b+rb')}\right)C_0'= J_{A_0},    \nonumber \\
C_0'' &-& \frac{2b\left( 3 + 10\frac{r^2}{\ell^2}  + 6f \right)}{3rf\left( 6b + rb' \right) } A_0' - \left(\frac{-\left( 4 + 10\frac{r^2}{\ell^2}  + 6f \right) }{2rf} + \frac{b'}{b} +  \frac{2b\left( 3 + 10\frac{r^2}{\ell^2}  + 6f \right) }{rf\left( 6b + rb' \right) }\right)C_0'= J_{C_0},
\eea
where $J_{A_0},\ J_{C_0}$ are source terms. Their expression contain about seventy terms so we don't give them here. Let us just mention that $J_{A_0},\ J_{C_0}$ are quadratic in $A_1,\ C_1$ and depend on the background functions.

Since equations \eqref{eqbr} depend only on the derivatives of $A_0$ and $C_0$, the solutions are invariant under
\be
A_0 \rightarrow A_0 + const.\ \ , \ \ C_0 \rightarrow C_0 + const.
\label{symbr}
\ee

\section{Near-horizon and asymptotics expansions}
\label{asymptotics}
The near-horizon behaviour for the background and for the first order is given in \cite{rms} (resp. \cite{rbd}), we will remind them. We also remind the asymptotic expansion for the background fields and compute it for the first order.

The near-horizon expansion for the background is given by \cite{rms}:
\be
\label{bghor}
a(r) = a_0 + a_0\frac{10r_h}{\left(5r_h^2 +  \ell^2\right)^2} (r-r_h) + \mathcal O(r-r_h)^2~,~
b(r) = b_1(r-r_h) + \mathcal O(r-r_h)^2~,~
f(r) =\frac{\left(2\ell^2+5r_h^2\right)}{r_h\ell^2}(r-r_h) + \mathcal O(r-r_h)^2,
\ee
where $a_0,\ b_1$ are numerical factors which are chosen in order to fit the following asymptotic expansion:
\bea
\label{bgas}
a(r)&=&\frac{r^2}{\ell^2}+\frac{2}{3} + \frac{1}{9}(\frac{\ell}{r})^2+c_z(\frac{\ell}{r})^3+\mathcal O(\ell/r^4)~,~
b(r)=\frac{r^2}{\ell^2}+\frac{2}{3} + \frac{1}{9}(\frac{\ell}{r})^2+c_t(\frac{\ell}{r})^3+\mathcal O(\ell/r^{4}),\nonumber\\
f(r)&=&\frac{r^2}{\ell^2}+ \frac{5}{6}+ \frac{2}{9}(\frac{\ell}{r})^2+(c_z+c_t)(\frac{\ell}{r})^3+ \mathcal O(\ell/r^{4}),
\eea
where $c_t,c_z$ are real constants to be determined numerically.

The near-horizon expansion for the first order is of the form \cite{rbd}:
\be
A_1(r) = A_{10} + A_{11}(r-r_h) + \mathcal O(r-r_h)^2~,~ C_1(r) = C_{10} + C_{11}(r-r_h) + \mathcal O(r-r_h)^2
\ee
where $A_{11},\ C_{11}$ are functions of $A_{10},\ C_{10}$ while the ratio $A_{10}/C_{10}$ must be determined numerically. Either $A_{10}$ or $C_{10}$ can be chosen arbitrarily since the equations are linear.

Using the asymptotic form \eqref{bgas} of the background, the first order fields have the following asymptotic expansions:
\bea
A_1(r)&=& -3\varphi\left(\frac{\ell}{r}\right)^5 + \frac{3\varphi
     \left(-4k^2\ell^2 + 48 \right)}{56}\left(\frac{\ell}{r}\right)^7 - 
  \frac{\varphi\left( 12k^4\ell^4 - 592k^2\ell^2 + 3137 \right) }
   {2016}\left(\frac{\ell}{r}\right)^9 + \mathcal O \left(\frac{\ell}{r}\right)^{11},\nonumber\\
C_1(r)&=& \varphi\left(\frac{\ell}{r}\right)^5 + \frac{\varphi \left( 12k^2\ell^2 - 136 \right)}{168}\left(\frac{\ell}{r}\right)^7 +  \frac{\varphi \left( 12k^4\ell^4 - 576k^2\ell^2 + 2921 \right) }{6048}\left(\frac{\ell}{r}\right)^9 + \mathcal O \left(\frac{\ell}{r}\right)^{11},
\eea
where $\varphi$ is a real constant determining the arbitrary normalisation of the fields.

Before considering the asymptotic solution for the backreacting fields, we give their near-horizon behaviour, since this expansion will be useful later:
\be
A_0(r) = A_{00} + A_{01}(r-r_h) + \frac{A_{02}}{2}(r-r_h)^2 + \mathcal O(r-r_h)^3~,~
C_0(r) = C_{00} + C_{01}(r-r_h) + \frac{C_{02}}{2}(r-r_h)^2 + \mathcal O(r-r_h)^3,
\ee
where $A_{00},\ C_{00},\ C_{01}$ are arbitrary real constants while 
\bea
A_{01} &=& -C_{01} - \frac{2\left( 4 + 10\frac{r_h^2}{\ell^2}  \right)\left( A_{10}^2 + 6A_{10}C_{10} - 3C_{10}^2 \right) k^2r_h}{3a_0}, \nonumber\\
A_{02} &=& \mathcal A_{02}(a_0,r_h,k,\ell,C_{01},A_{10},C_{10})~,~
C_{02} = \mathcal C_{02}(a_0,r_h,k,\ell,C_{01},A_{10},C_{10}).\nonumber
\eea
The functions $\mathcal A_{02},\ \mathcal C_{02}$ are rather cumbersome and we refrain to give them explicitely. Here, we posed $a_0 \equiv a(r_h),\ A_{10} \equiv A1(r_h),\ C_{10} \equiv C_1(r_h),\ C_{01}\equiv C'(r_h)$.

The most general asymptotic expansion for the back reaction fields is given by:
\bea\label{asympbr}
A_0(r) &=& \alpha_0 - \frac{5\left( c_t - c_z\right) \sigma }{4}\left(\frac{\ell}{r}\right)^6 + \frac{-480\chi  + \sigma}{160}\left(\frac{\ell}{r}\right)^5 + \left[\frac{65\chi}{28} - \frac{13\sigma}{2688}\right]\left(\frac{\ell}{r}\right)^7 + \mathcal O\left( \frac{\ell}{r}\right)^8,\\
C_0(r)&=&\varphi_0 + \left[\frac{\ell}{r} -\frac{1}{36}\left(\frac{\ell}{r}\right)^3\right]\sigma + \chi \left(\frac{\ell}{r}\right)^5 +      \frac{\left( 4c_t - 5c_z \right) \sigma}{12}\left(\frac{\ell}{r}\right)^6    
   +\left[\frac{119\sigma }{36288} -  \frac{155\chi}{252} \right]\left(\frac{\ell}{r}\right)^7 + \mathcal O\left( \frac{\ell}{r}\right)^8,\nonumber
\eea
with $\alpha_0,\ \varphi_0$ are arbitrary real constants and $\chi,\ \sigma$ are real constants to be determined numerically.

Properties of the solution will be discussed in section \ref{properties}. Some of these properties depends on the asymptotic of the solution. One should mention herethat the term proportional to $1/r$ in \eqref{asympbr} gives a diverging contribution to the mass and is thus unphysical. Also, the constants $\alpha_0,\ \varphi_0$ must be chosen as zero, since we are dealing with perturbations. For solutions with $\sigma=\alpha_0=\varphi_0 = 0$, the asymptotic form \eqref{asympbr} simplifies considerably and gives to the leading order~:
\be
A_0(r) = -3\chi\left( \frac{\ell}{r}\right)^5 + \mathcal O\left( \frac{\ell}{r}\right)^7\ \ , \ \  
C_0(r) = \chi\left( \frac{\ell}{r}\right)^5 + \mathcal O\left( \frac{\ell}{r}\right)^7
\label{asphy}
\ee

Let us notice that the equations of perturbations, at each order in $\epsilon$ and each massive modes are basically equations for waves propagating in the background of $a,\ b,\ f$ with a source terms. Let us write the perturbative expansion as $X(r,z) = \sum_i \sum_j X_i^j(r)\cos(ikz) \epsilon^j$, $X$ denoting the functions $A,\ B, C$. Then, for a given mode $X_i^j$, the equation is of the form
\be
\Box X_i^j = J_{X_i^j}(X_m^{n<i}),
\ee
where $\Box$ is the d'Alembertian for the background and $J_{X_i^j}$ is a polynomial of the $X_m^n$. The exponent of each term in the polynomial times the order in $\epsilon$ of the term must be equal to $j$ in order to enter the equation at the $\epsilon^j$ order. 

The first mode containing a source term is the zero mode and decays as $1/r^5$. The mode $X_1^1$ also decays as $1/r^5$, thus source terms for the order $3$ in $\epsilon$ will be of the form $(X_1^1)^3,\ X_0^2 X_1^1$ and decay as $1/r^{15}$. This decay is too quick to influence the leading order in the asymptotic expansion, thus leaving the free wave decay $1/r^5$. This argument can be extended by recurrence.

\section{Boundary Conditions and Numerical Techniques}
\label{num}

The background and first order boundary conditions are reminded here: 
\be
a(r_h) = a_0 \ ,\ f(r_h) = 0 \ ,\ b(r_h) = 0\ , \ b'(r_h) = b_1,
\ee
where $a_0,\ b_1$ are real constants chosen such that the asymptotic solution  fits the Feffer-Graham expansion \eqref{bgas}.
For the first order fields, we have
\be
\label{bco1}
A_1(\infty)= 0\ ,\ C_1(\infty) = 0 ~,~ A_1'(r_h) = \mathcal{A_1}(A_{10},C_{10},a_0,b_1) ~,~ C_1'(r_h)   = \mathcal{C_1}(A_{10},C_{10},a_0,b_1),
\ee
where $\mathcal{A_1},\ \mathcal{C_1}$ are the nonlinear conditions ensuring that the right hand side of \eqref{eqF1} are regular at the horizon. The four conditions \eqref{bco1} are not sufficient to fix the normalisation of $A_1,\ C_1$. In order to solve this problem we add an extra equation expressing the constancy of the eigen value, $(k^2)' = 0$, in order to solve the problem \cite{rbd} and take advantage of the supplementary freedom to fix the arbitrary normalisation of $F_1$ by means of $C_1(r_h) = 1$.

For the backreacting fields, the equations imply only one regularity condition for the equation of $A_0'(r_h)$, and the perturbations $A_0,C_0$ must vanish at infinity. 
\bea\label{bcbr}
A_0(\infty) = 0\ ,\ C_0(\infty) = 0 ~,~A_0'(r_h) = \mathcal{A}_0(C_0'(r_h),A_{10},C_{10},a_0,b_1),
\eea
where $\mathcal{A}_0$ is the regularity condition for $A_0$.
This leaves a free parameter, for example $C_0'(r_h)$, which is fixed imposing the physical decay \eqref{asphy}.

\par In practice, we integrate the background and first order with the solver \emph{Colsys}, based on a Newton Raphson algorithm, with a mesh adapted in order to minimise the error \cite{COLSYS}. The backreacting fields tends naturally to constants for large values of $r$. With the invariance \eqref{symbr}, we give arbitrary values to $A_0(r_h),\ C_0(r_h)$, set $A_0'(r_h)$ according to the regularity condition and shoot using $C_0'(r_h)$ as a shooting parameter in order to fulfil the decay condition \eqref{asphy}, \emph{transposed to the derivative} of $C_0(r)$.

We used a Runge Kutta algorithm at order $4$ to integrate the backreacting fields. The consistency of the result has been checked by comparing the result of the numerical integration with the analytic result of the asymptotic expansion.

\section{Properties of the solution}
\label{properties}
\subsection{Scaling properties and deformation parameter}

The equations present the following symmetry:
\bea
\label{scal}
\ell\rightarrow \alpha \ell\ \ , \ \ r_h \rightarrow  \alpha r_h \ \ , \ \ k\rightarrow k/\alpha,
\eea
for real values of $\alpha$.
Thus, given solutions with fixed AdS radius and varying horizon radius, one can convert these solution to fixed radius and varying AdS radius. This is usually referred to as copies of solutions. 

The scaling \eqref{scal} affects the thermodynamical quantities in the following way~:
\bea
S \rightarrow \alpha^{3} S \ \ ,\ \ T_H \rightarrow T_H/\alpha \ , \ \mathcal T \rightarrow  \alpha^2 \mathcal T \ \ ,\ \ M\rightarrow \alpha^2 M 
\eea

Note that, in opposition to the asymptotically flat case, there is an intrinsic length scale in the model, given by $\ell$. As already mentioned in \cite{rbd}, the solutions are characterised by two dimensionless quantities, $\mu_1,\ \mu_2$, with
\be
\mu_1 = \frac{M}{16\pi G L^{d-3}}\ \ , \ \ \mu_2 = \frac{L}{\ell}
\label{mu12}
\ee
with $M$ the mass of the black string.

Following \cite{gubser}, one defines a deformation parameter, which quantifies how much the solution deviates from the uniform black string case. This parameter is given by:
\be
\lambda = \frac{R_{\mbox{max}}}{R_{\mbox{min}}} - 1
\ee
where $R_{\mbox{max}},\ R_{\mbox{min}}$ represents the minimal and maximal areal radius at the horizon.
Since we are dealing with \textit{perturbative} non uniform black strings, our results will be valid for $\lambda << 1$. The areal radius at the horizon is given by:
\be
R(z) = \left.\sqrt{g_{\theta\theta}}\right|_{r=r_h} = r_h e^{C(r_h,z)}.
\ee
In the case of our interest, $R_{\mbox{max}} = R(0)\ , \ R_{\mbox{max}} = R(\pi)$ which leads to:
\be
\lambda = 2C_{10}\epsilon\left(1 + \epsilon C_{10}  \right) + \mathcal O(\epsilon)^3
\ee


\subsection{Global charges and thermodynamical quantities}
The conserved charges of the solutions are computed using the counterterms approach \cite{counter}. This procedure, applied to uniform warped AdS black strings is explained in details in \cite{rms}. In this approach, one has to add suitable boundary terms $S_{ct}(h_{\mu\nu})$ to the action, in order to cancel the diverging terms, coming from the contribution $AdS$ background. These terms do not contribute to the equations of motion and can be found in \cite{counter}.

The variation of the total action $S_{tot} = S + S_{ct}$ with respect to the boundary metric $h_{\mu\nu}$ then leads to a regularized boundary stress tensor $T_{\mu\nu} = \frac{2}{\sqrt{-\gamma}}\frac{\delta S_{tot}}{\delta h^{\mu\nu}}$. Then, a conserved charge
\be
\label{gcharges}
\mathcal Q_{\hat\xi} = \int_\Sigma \hat\xi^a T_{ab} n^b dS_4
\ee
can be associated with a closed four-surface $\Sigma$ with surface element $dS_4$ and normal $n^b$, provided $\hat\xi^a$ is a Killing vector of the boundary geometry.

The first relevant corrections on thermodynamical quantities arises from the back reaction, since the integral \eqref{gcharges} involves an integration over $z$. However, terms with $\cos(k z),\ \cos(2k z)$ factors cancel once integrated on $z\in[0,L],\ L=\frac{2\pi}{k}$. 
A straightforward computation gives the mass of the solution, which is the charge associated to the Killing vector $\partial_t$:
\be
M = \frac{3\ell^2L\pi}{64 G}\left( -4c_t + c_z + 30\epsilon^2 \chi  \right).
\ee
However, the solution's tension, which is the charge associated with $\partial_z$ is not affected by the perturbation, leaving the same value for the pNUBS than for the background:
\be
\mathcal T = \frac{3\ell^2\pi}{64 G}\left( -4c_z + c_t \right).
\ee
Let us define also the relative tension $n = \mathcal{T}L/M$. In the perturbative approach, these quantities can be expanded in powers of $\epsilon$; i.e. keeping the first correction $\bar n + \epsilon^2\delta n\ ,\ \bar\mu_i +\epsilon^2\delta\mu_i,\ i=1,2$, with the bar referring to the quantities evaluated on the background solution. After some algebra, we find the following relations involving the dimensionless quantities \eqref{mu12}:
\be
\delta\mu_1 = 30\epsilon^2\chi \ \ ,\ \ \delta\mu_2 = 0 \ \ ,\ \ \frac{\delta n}{n} =-\frac{\delta\mu_1}{\mu_1}.
\ee

The entropy of the pNUBS, which is one quarter of the event horizon area, is given by
\be
S =  2\sqrt{a_0}L\pi^2r_h^3\left(1 - 
    \frac{\left( 5a_0\left( A_{10}^2 - 12{F_{00}} - 6{A_{10}}{F_{10}} - 9{{F_{10}}}^2 \right)  + 
         3\left( {A_{10}} - {F_{10}} \right) {F_{10}}k^2\ell^2 \right) }{10
      a_0} {\epsilon }^2\right).
\ee
Depending on the sign of the correction, the pNUBS and thus presumably the full NUBS phase may be thermodynamically preferred rather than the uniform black string.

The Hawking temperature as found by demanding regularity on the Euclidean section near the horizon, is given by

\be
T_H = \sqrt{\frac{b_1}{r_h\ell^2}\left(2\ell^2+5r_h^2\right)}\left( 1 + \epsilon^2 \frac{10a_0\left( 2A_{00} + A_{10}^2 \right)  + 3\left( A_{10} - F_{10} \right) F_{10}k^2\ell^2}
  {20a_0} \right).
\ee

In $6$ dimensions, the stress tensor has vanishing trace at the background level \cite{rms}. Let us finally mention that, as expected, this property holds for perturbative non uniform black strings.

\section{Numerical results}
\label{results}
In the figures we present, we use the rescaling \eqref{scal} to choose the AdS radius to be $\ell= \sqrt{10}$ without loosing generality. This value of $\ell$ corresponds to a unit negative cosmological constant.

\begin{figure}[H]
	\centering
		\includegraphics[scale=.7]{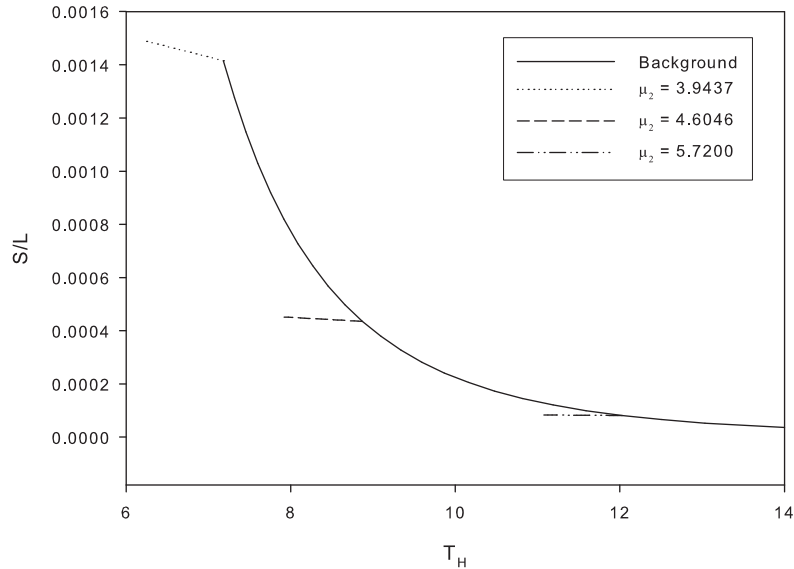}
	\caption{The entropy per unit length as a function of the temperature for various $\mu_2$. The entropy grows while the temperature decreases; for larger values of $\mu_2$, the entropy is nearly constant.}
	\label{fig:ST}
\end{figure}

As already mentioned, the solutions are characterized by two dimensionless quantities, $\mu_1,\ \mu_2$. In fig. \ref{fig:ST}, we give the entropy per unit length as a function of the horizon temperature for the uniform black strings and the first relevant corrections for perturbative non uniform black strings, i.e. when the deformation parameter $\lambda<<1$, for different values of $\mu_2$. Since large black strings are known to be stable, we consider perturbative non uniform small black strings. Actually, the entropy per unit length of the background (the uniform phase) does not depend on the length of the black string. So this diagram is to be understood in the following way: each intersection of the non uniform black string phases with the uniform string case is the emanating point of a phase transition at fixed rescaled extradimension length $\mu_2$. 

For every cases we considered, the entropy of the non uniform phase is higher than the entropy of the uniform phase for the corresponding $\mu_2$ and the temperature decreases. For larger $\mu_2$, the entropy still increasing in the non uniform phase, but less than when $\mu_2$ is small. This can be understood in the following way: since the length becomes of the order of $\ell$, so this gives "small - long black strings"; small for $r_h$, large for $L$ and the effect of the $AdS$ radius is felt by the string in the $z$-direction. This is a consequence of the fact that the solution is characterised by two dimensionless constants.

\begin{figure}[H]
	\centering
		\includegraphics[scale=.7]{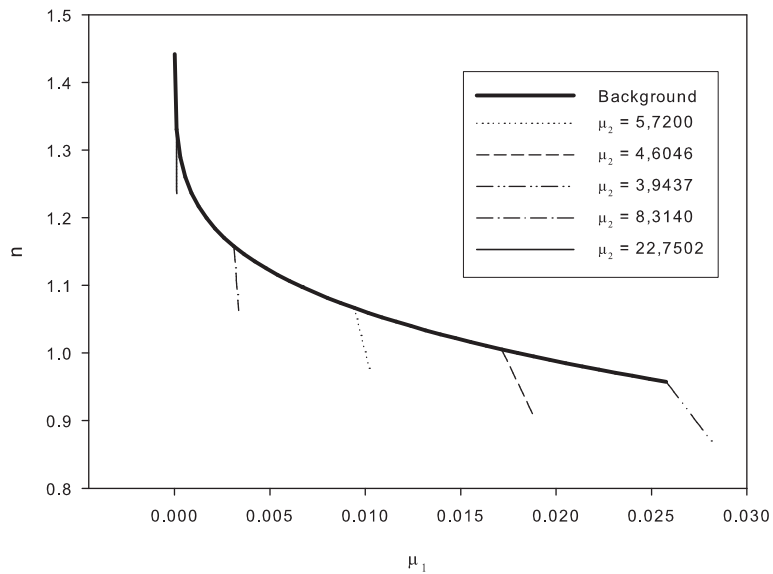}
	\caption{The relative tension as a function of the rescaled mass for various rescaled length. }
	\label{fig:nMu}
\end{figure}

Figure \ref{fig:nMu} is a plot of the relative tension $n$ as a function of the rescaled mass $\mu_1$. This diagram is drawn in the same spirit as figure \ref{fig:ST}. The relative tension of the non uniform phase is smaller than the uniform phase, but the mass is higher.

As it should have been expected, the flat space picture seems to hold for $AdS$, at least for small-short pNUBS: non uniform black strings is preferred when $r_0\approx L$.

\section{Conclusion}
\label{end}

We constructed the first corrections to the thermodynamical quantities of AdS non uniform black strings in six dimensions. The main features of the phase diagram is similar to the asymptotically flat space case, except that the relative tension now depends on the rescaled mass for fixed rescaled length ($n = n(\mu_1)\mbox{, for fixed }\mu_2 $).  

For the range of parameters we considered, the correction to the entropy was always positive, althought nearly vanishing for large rescaled length. The correction on the temperation was always negative, implying that the non uniform black string phase with small deformation parameter near the emanating point should be colder and more entropic, thus thermodynamically preferred. The tension-mass ration is decreasing in the non uniform phase, while the rescaled mass is increasing. It is remarkable that the tension is unaffected by the non uniformity, at least at second order in perturbation theory.

It seems likely that small-short stable uniform black string should decay to caged AdS black holes, as it is the case in asymptotically flat space \cite{kol} while the unstable branch should decay to non uniform black string. We saw that the amount of change in entropy per unit length decreases when the rescaled length increases, but is still higher than the uniform phase. It was already known that the ratio $r_h/\ell$ plays an important role in the problem, but it seems that so does the rescaled length $\mu_2$. It seems likely that the string is affected by the AdS when the rescaled length is large i.e. when the length is of the same order than the AdS radius.

Some interesting outlooks for this work should be to construct the full nonlinear solution for the warped AdS non uniform black string and the connection with the thermodynamics of the boundary CFT. Moreover, it would be interesting to generalise this work to $d$ dimensions; in asymptotically flat space, there exists a critical dimension above which where the entropy of the non uniform branch increases instead of decreasing \cite{sorkin}. The consequence is that in $d\leq13$ the uniform branch is more stable than the non uniform branch.

\section{Acknowledgement}
I wish to gratefuly acknowledge Y. Brihaye for usefull discussions and advices.



\end{document}